\documentclass[referee]{aa} 
\input{psfig.tex}
\def\gsim{\;\lower.6ex\hbox{$\sim$}\kern-7.75pt\raise.65ex\hbox{$>$}\;}
\def\lsim{\;\lower.6ex\hbox{$\sim$}\kern-7.75pt\raise.65ex\hbox{$<$}\;}
\begin{document}

%

%
\title{ The O-Na and Mg-Al Anticorrelations in Turn-Off and early Subgiants in
Globular Clusters\footnote{Based on data collected at the European Southern
Observatory, Chile} }

\author{
R.G. Gratton\inst{1}, 
P. Bonifacio\inst{2},
A. Bragaglia\inst{3}, 
E. Carretta\inst{1}, 
V. Castellani\inst{4},
M. Centurion\inst{2}, 
A. Chieffi\inst{5},
R. Claudi\inst{1}, 
G. Clementini\inst{3},
F. D'Antona\inst{6},
S. Desidera\inst{1,7},
P. Fran\c cois\inst{8},
F. Grundahl\inst{9},
S. Lucatello\inst{1,7},
P. Molaro\inst{2},
L. Pasquini\inst{8}, 
C. Sneden\inst{10},
F. Spite\inst{11},
\and
O. Straniero\inst{12}
}

\authorrunning{R.G. Gratton et al.}
\titlerunning{O-Na anticorrelations in TO stars in globular clusters}


\offprints{R.G. Gratton}

\institute{
Osservatorio Astronomico di Padova, Vicolo dell'Osservatorio 5, 35122
 Padova, Italy\\
\and
Osservatorio Astronomico di Trieste, Italy\\
\and
Osservatorio Astronomico di Bologna, via Ranzani 1, 40127 Bologna, Italy\\
\and
Dipartimento di Fisica, Universit\`a di Pisa, Italy\\
\and
Istituto di Astrofisica Spaziale, CNR, Italy\\
\and
Osservatorio Astronomico di Roma, Italy\\
\and
Dipartimento di Astronomia, Universit\`a di Padova, Italy\\
\and
European Southern Observatory\\
\and
Department of Astronomy, University of Aarhus, Denmark\\
\and
The University of Texas at Austin, USA\\
\and
Observatoire de Meudon, France\\
\and
Osservatorio Astronomico di Collurania, Italy
}

\date{Received ; accepted }

\abstract{
High dispersion spectra ($R\gsim 40,000$) for a quite large number of stars at
the main sequence turn-off and at the base of the subgiant branch in NGC6397
and NGC6752 were obtained with the UVES on Kueyen (VLT UT2). The [Fe/H] values
we found are $-2.03\pm 0.02\pm 0.04$\ and $-1.42\pm 0.02\pm 0.04$\ for NGC6397
and NGC6752 respectively, where the first error bars refer to internal 
and the second ones to systematic errors. In both clusters the [Fe/H]'s obtained
for TO-stars agree perfectly (within a few per cents) with that obtained for
stars at the base of the RGB. The [O/Fe]=$0.21\pm 0.05$\ value we obtain for
NGC6397 is quite low, but it conforms to previous results obtained for giants
in this cluster. Moreover, the star-to-star scatter in both O and Fe is very
small, indicating that this small mass cluster is chemically very homogenous.
On the other side, our results show clearly and for the first time that the
O-Na anticorrelation (up to now seen only for stars on the red giant branches
of globular clusters) is present among unevolved stars in the globular cluster
NGC6752, a more massive cluster than NGC6397. A similar anticorrelation is
present also for Mg and Al, and C and N. It is very difficult to explain the
observed Na-O, and Mg-Al anticorrelation in NGC6752 stars by a deep mixing
scenario; we think it requires some non internal mechanism.
}



\maketitle

\section{INTRODUCTION}

Globular clusters (GCs) are (with a few significant exceptions, most notably
the case of $\omega$~Cen) generally very homogenous as far as Fe-peak elements
are considered; on the other side, abundances of the lighter elements (from C
to Al) show a complex, not yet adequately explained pattern. Rather wide
variations in the strength of the CH and CN bands among stars on the red giant
branch (RGB) were already noticed in the seventies (Osborn 1971, and many
other references; for reviews see Smith 1987, Kraft 1994), and found to be
correlated with other cluster properties (Norris 1987). More recently, a CN-CH
anticorrelation was found in Main Sequence (MS) stars of NGC~6752 (Suntzeff \&
Smith 1991), 47~Tuc (Cannon et al. 1998), and M~71 (Cohen 1999a) from low
dispersion spectra. Even more striking is the Na-O anticorrelation among GC
red giants discovered by the Lick-Texas group (see e.g. Kraft et al. 1997).

These features are indications of the presence of elements processed through
the complete CNO-cycle in the atmospheres of some globular cluster stars; at
temperatures where complete CNO cycle occurs, quite large amounts of Na may be
produced by proton capture on $^{22}$Ne nuclei (Denissenkov \& Denissenkova
1990; Langer \& Hoffman 1995; Cavallo, Sweigart \& Bell 1996). Whatever the
cause of these abundance anomalies, it appears to be related to the dense
environment of GCs. In fact, field halo stars present surface abundances of Li,
C, N, O, Na that are well explained by standard models, once a second mixing
episode (only affecting Li, C and N abundances) after the RGB-bump is allowed
(Gratton et al. 2000). Note that the RGB bump plays an important r\^ole,
because there the molecular weight barrier created by the deepest inward
penetration of the outer convective envelope (at the base of the RGB) is
canceled by the outward shift of the H-burning shell. The absence of this
barrier in the following evolution allows deep mixing caused by e.g.
meridional circulations (Sweigart \& Mengel 1979; Charbonnel 1994; Denissenkov
\& Tout 2000). Hence, both the O-Na anticorrelation observed among giants,
and the C and N anomalies observed in stars in different evolutionary phases,
are peculiar to GCs.

The proposed explanations of the abundance anomalies in globular cluster stars
have been oscillating in years between an "in situ" mechanism (very deep mixing
in RGB stars, possibly activated by a core rotation that might be larger in
cluster than in field stars) and an external source of material (either
primordial proto-cluster gas or processed material from a polluting
companion). Since MS stars do not have a convective envelope able to enhance N
while depleting C, these anomalies should be attributed mainly to a primordial
origin and/or to pollution, while evolutionary mixing could possibly be
responsible only for likely minor effects in RGB stars. However, rather
surprisingly, an analysis of stars below the Turn-off (TO) of M~13 (Cohen
1999b) shows no variations in the strength of CN and CH molecular features,
likely excluding primordial variation of C and N abundance in this cluster,
which instead presents the clearest Na-O anticorrelation on the RGB (Kraft et
al. 1997), and a range in the strength of CN and CH bands among giants
(Suntzeff 1981). While the result of Cohen is very surprising, the analysis is
only qualitative (a quantitative analysis being planned in a future, still
unpublished paper), so that it is not clear which is the upper limit found to
C and N abundance variations among MS stars in this very important cluster.
In general, the relation between the Na-O and C-N anticorrelation is not yet
well established. Observation of a Na-O anticorrelation among main sequence
stars of globular clusters would be crucial, clearly ruling out the deep
mixing scenario.

Earlier explorative studies of stars close to the TO and at the base of the
subgiant branch of a few clusters have been presented by King, Stephens \&
Boesgaard (1998), Deliyannis, Boesgaard \& King (1995) and Boesgaard et al.
(1998) who used HIRES at Keck to gather spectra of three stars in M92; and
Pasquini \& Molaro (1996, 1997) who exploited the presence of closer clusters
in the southern emisphere to obtain spectra of a few stars in NGC6397 and 47
Tuc using EMMI at NTT (these stars are about a magnitude brighter than those of
M92). However, only a handful of stars could be observed, and at very low S/N.
Only UVES (D'Odorico et al. 2000) at the Kueyen (VLT UT2) Telescope is able to
provide abundances of O, Na, Mg, Al and other elements for significant samples
of well studied stars at the TO and at the base of RGB (well below the RGB
bump), crucial to clarify most of these issues. Here we present a short
description of the first results of the ESO Large Program 165.L-0263 devoted to
acquisition and analysis of high dispersion spectra ($R\gsim 40,000$) of a
rather large sample of stars at the TO and at the base of the RGB in NGC6752,
NGC6397, and 47 Tuc. A more detailed presentation of the analysis methods, and
of related results for Li, will be given in forthcoming papers, still in
preparation.

\begin{figure*} 
\psfig{figure=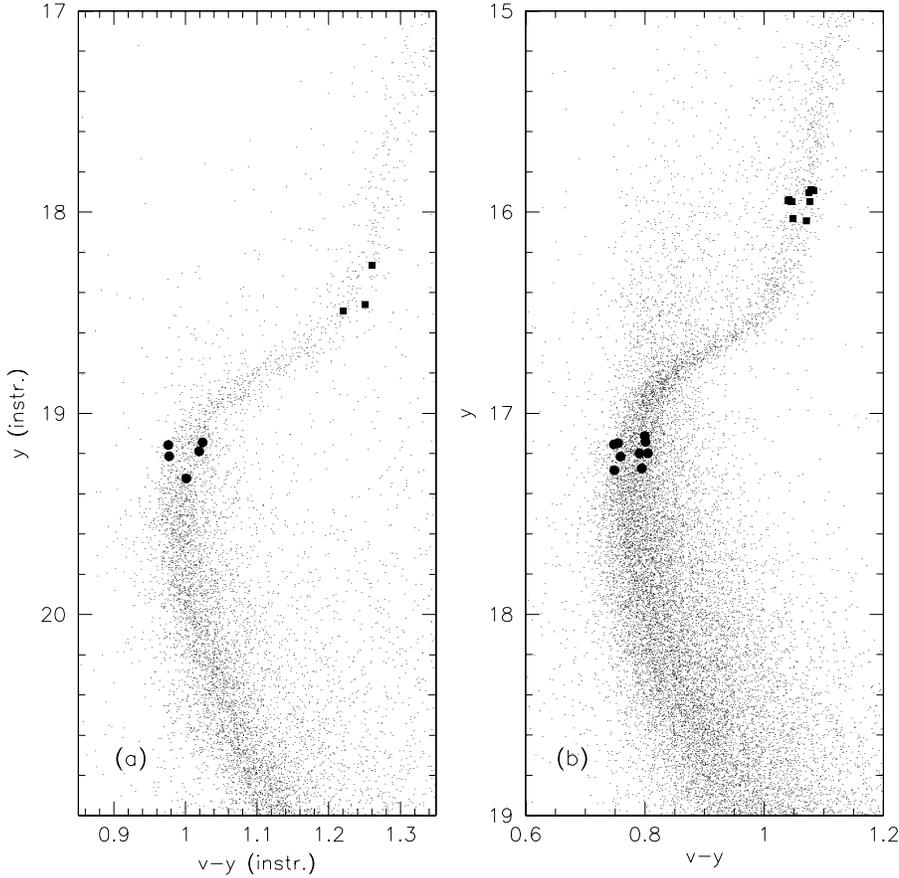,width=13.0cm,clip=}
\caption[]{Positions of program stars (large symbols) on the Str\"omgren
colour-magnitude diagram of NGC6397 (panel a) and NGC6752 (panel b). Photometry
is from Grundahl et al. (1999). Note that photometry for NGC6397 is still
uncalibrated, so that only instrumental magnitudes are given}
\label{f:fig1} 
\end{figure*}  

\begin{figure} 
\psfig{figure=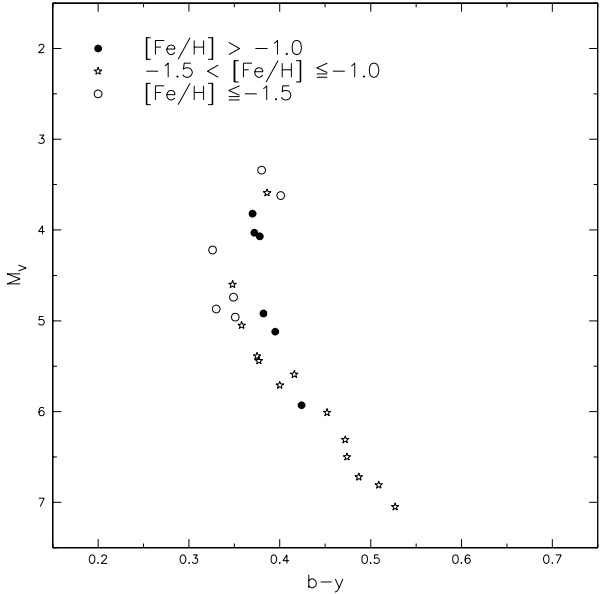,width=8.8cm,clip=}
\caption[]{Colour-magnitude  diagram for the field stars; M$_{\rm V}$ is
computed using Hipparcos parallaxes}
\label{f:fig2} 
\end{figure}  

\section{SAMPLE SELECTION AND OBSERVATIONS}

Observations were carried out in two 6-nights runs (June and September 2000);
due to poor weather conditions in the September run, we were not able to
complete the acquisition of the spectra in 47 Tuc. Hence the present paper
deals only with data for the two other clusters. In the same runs we also
acquired data for 25 field stars with good parallaxes from the Hipparcos
catalogue and metallicity similar to that of the GCs.

For each cluster, we selected two groups of stars: (i) dwarfs about 0.2 mag
brighter than the turn-off (5 stars in NGC6397 and 9 in NGC6752); and (ii)
stars at the base of the red giant branch (3 stars in NGC6397 and 9 stars in
NGC6752), hereinafter called subgiants. Positions of the observed stars in the
c-m diagrams of the two clusters (from Grundahl et al. 1999) are given in
Figure~\ref{f:fig1}. A posteriori, radial velocities and chemical composition
show that all stars are bona fide members of the clusters. Stars with likely
strong and weak CN bands were selected from Str\"omgren photometry: the $c_1$
index seems in fact to be higher for CN-richer stars (see Grundahl et al.
1999). Given this choice, it should be clear that stars are not a random
sample of stars in the different magnitude bins, but rather they were selected
in order to include quite extreme cases of CN-poor and CN-rich stars. For
comparison, we give in Figure~\ref{f:fig2} also the location in a similar c-m
diagram of the field stars we observed.

The observations were carried out using the dichroic beamsplitter \#2. This
allowed to observe simultaneously two spectral ranges with the two arms of the
spectrograph. Slightly different setups were used in the two runs: the blue
spectra cover the spectral range from about 3,500~\AA\ to about 4,700~\AA,
while the red spectra cover the range from 5,700~\AA\ to about 8,700~\AA. This
very broad spectral range allows observations of a large number of spectral
features. The resolution was set by the slit width: in most cases this was 
1~arcsec (yielding a resolution of 43,000 at order centers); however, this
value was sometimes changed according to the variable seeing conditions, from
0.7~arcsec in the best cases, down to 1.2~arcsec in the worst ones. The slit
length was always set at 8~arcsec, allowing accurate sky subtraction (necessary,
since observations were carried out in full moon). Most of the objects are
very faint ($V\gsim 17$), so that long exposures were required even with an 8~m
telescope like Kueyen, and a very efficient spectrograph like UVES. Typical
exposure times were $\sim 1$~hr for subgiants, and about 4 hr (splitted in 3
to 4 segments) for TO stars. The S/N (per pixel at 6700~\AA; there are about 5
pixels per resolution element) of the resulting spectra is $\sim 100$\ in the
best cases, and down to about 20 in the worst ones. Relevant data are
presented in Tables~1, 2, and 3.

\begin{table*}
\caption{Parameters for field stars}
\begin{tabular}{rrrrrrrrr}
\hline
   Star & $M_v$ &$B-V$ &$b-y$ &[Fe/H] &$\log g$&$T_{\rm eff}$&$T_{\rm eff}$& 
$T_{\rm eff}$\\
HD/BD   &     &      &      &        &      &H$_\alpha$&Best&Alonso et al.\\
\hline
   10607 &4.03 &0.56  &0.372 &$-$0.95 & 4.00 & 5780 & 5734 &        \\
   29907 &6.01 &0.63  &0.452 &$-$1.47 & 4.82 & 5636 & 5406 &        \\
   31128 &5.05 &0.49  &0.358 &$-$1.44 & 4.63 & 6279 & 5969 &        \\
  108177 &4.87 &0.43  &0.330 &$-$1.70 & 4.44 & 5884 & 6063 &   6067 \\
  116064 &4.74 &0.45  &0.349 &$-$1.87 & 4.35 & 5770 & 5923 &        \\
  120559 &5.93 &0.66  &0.424 &$-$0.90 & 4.51 & 5207 & 5378 &        \\
  121004 &5.12 &0.62  &0.424 &$-$0.75 & 4.32 & 5401 & 5548 &        \\
  126681 &5.71 &0.60  &0.400 &$-$1.13 & 4.55 & 5444 & 5556 &   5541 \\
  132475 &3.62 &0.56  &0.401 &$-$1.61 & 3.79 & 5596 & 5520 &   5788 \\
  134169 &3.82 &0.565 &0.370 &$-$0.81 & 4.06 & 6106 & 5825 &   5874 \\
  134439 &6.72 &0.77  &0.487 &$-$1.31 & 4.78 & 5151 & 5052 &   4974 \\
  134440 &7.05 &0.85  &0.527 &$-$1.28 & 4.69 & 4777 & 4777 &   4746 \\
  140283 &3.34 &0.49  &0.380 &$-$2.46 & 3.67 & 5560 & 5645 &   5691 \\
  145417 &6.81 &0.82  &0.509 &$-$1.23 & 4.78 & 5096 & 4922 &        \\
  159482 &4.92 &0.58  &0.382 &$-$0.83 & 4.32 & 5546 & 5656 &        \\
  166913 &4.22 &0.45  &0.326 &$-$1.59 & 4.08 & 5921 & 6020 &        \\
  181743 &4.96 &0.45  &0.351 &$-$1.81 & 4.42 & 5749 & 5927 &        \\
  188510 &5.59 &0.60  &0.416 &$-$1.44 & 4.50 & 5412 & 5485 &   5564 \\
  189558 &3.59 &0.575 &0.386 &$-$1.15 & 3.87 & 5829 & 5670 &   5663 \\
  193901 &5.44 &0.55  &0.377 &$-$1.03 & 4.62 & 5848 & 5780 &   5750 \\
  194598 &4.60 &0.49  &0.348 &$-$1.15 & 4.42 & 6064 & 5980 &   6017 \\
  204155 &4.07 &0.57  &0.378 &$-$0.70 & 4.06 & 5907 & 5751 &        \\
  205650 &5.39 &0.51  &0.375 &$-$1.12 & 4.52 & 5628 & 5782 &        \\
-35~0360 &6.31 &0.76  &0.472 &$-$1.08 & 4.52 & 4957 & 5086 &   4980 \\
+05~3640 &6.50 &0.74  &0.474 &$-$1.15 & 4.62 & 4987 & 5063 &        \\
\hline
\end{tabular}
\end{table*}

\begin{table*}
\caption[]{Data for NGC~6397 stars (identifications from Grundahl et al. 1999);
Al abundances from resonance lines}
\begin{tabular}{rrrrrrrrrrrrr}
\hline
Star   &$S/N$&n.& [Fe/H] &r.m.s&EW & [O/Fe]&[O/Fe] & EW & EW & [Na/Fe]
&[Mg/Fe] & [Al/Fe]\\
       &    &lines&      &     &7771& LTE  & nLTE  &8184&8190& nLTE  \\
\hline
\multicolumn{13}{c}{Dwarfs}         \\
  1543 & 91 & 23 &$-$2.02 & 0.12 & 14.7 &   0.32 &   0.16 &  ..  & 38.8 & 0.21
& 0.06   & $-$0.40 \\ 
  1622 & 82 & 21 &$-$2.02 & 0.10 & 13.3 &   0.27 &   0.11 & 28.5 & 40.7 & 0.28
& 0.10   & $-$0.34 \\ 
  1905 & 92 & 22 &$-$2.06 & 0.11 & 13.4 &   0.27 &   0.11 & 17.1 & 29.4 & 0.02
& 0.12   & $-$0.35 \\
201432 & 97 & 22 &$-$2.00 & 0.11 & 12.8 &   0.25 &   0.08 & 17.0 & 35.6 & 0.08
& 0.06   & $-$0.50 \\ 
202765 & 59 & 18 &$-$2.02 & 0.14 & 16.1 &   0.38 &   0.21 & 20.7 & 28.5 & 0.06
&$-$0.06 & $-$0.41 \\ 
\multicolumn{13}{c}{Subgiants}      \\
   669 & 91 & 27 &$-$2.01 & 0.11 & ~6.0 &   0.34 &   0.26 & 62.9 & 86.8 & 0.48
& 0.23   & $-$0.06 \\ 
   793 &105 & 31 &$-$2.04 & 0.12 &$<$6.0&$<$0.34 &$<$0.26 & 39.9 & 72.1 & 0.21
& 0.28   & $-$0.21 \\ 
206810 & 85 & 23 &$-$2.10 & 0.10 & ~8.0 &   0.48 &   0.48 & 49.5 & 68.3 & 0.25
& 0.10   & $-$0.24 \\ 
\hline
\end{tabular}
\end{table*}

\begin{table*}
\caption[]{Data for NGC~6752 stars. Stars are ordered according to increasing
Na abundance. Star identifications are from Grundahl et al. (1999). Note: 1) Al
abundances from resonance lines; (2) Al abundances from high excitation
8772.9/73.9 doublet}
\begin{tabular}{rrrrrrrrrrrrr}
\hline
Star   &$S/N$&n.& [Fe/H] &r.m.s&EW & [O/Fe]&[O/Fe] & EW & EW & [Na/Fe]
&[Mg/Fe] & [Al/Fe]\\
       &    &lines&      &     &7771& LTE  & nLTE  &8184&8190&  nLTE  \\
\hline
\multicolumn{13}{c}{Dwarfs}         \\
  4428 & 49 & 22 &$-1.52$ & 0.14 &  35.7 &  0.35 &   0.24 & 24.4 & 57.1 
& -0.35 &   0.04 & $-$1.07~(1) \\
  4383 & 49 & 23 &$-1.42$ & 0.19 &  46.0 &  0.60 &   0.49 & 33.6 & 61.5 
& -0.23 &   0.01 & $-$0.70~(1) \\
202316 & 44 & 21 &$-1.56$ & 0.20 &  28.7 &  0.31 &   0.20 & 42.6 & 69.0 
& -0.09 &$-$0.06 & $-$1.04~(1) \\
  4341 & 42 & 23 &$-1.57$ & 0.26 &  21.6 &  0.30 &   0.21 & 63.4 & 83.7 
&  0.18 &   0.06 & $-$1.35~(1) \\
  4458 & 33 & 17 &$-1.52$ & 0.27 &  19.4 &  0.04 &$-$0.06 & 71.9 & 82.6 
&  0.24 &$-$0.09 & $-$1.22~(1) \\
  4661 & 21 & 16 &$-1.24$ & 0.17 &   ..  &   ..  &    ..  & 75.4 & 84.3 
&  0.28 &   0.15 & $-$0.68~(1) \\
  5048 & 44 & 25 &$-1.43$ & 0.12 &  11.0 &$-$0.29&$-$0.37 & 78.2 & 93.7 
&  0.37 &$-$0.24 & $-$0.40~(1) \\
  4907 & 66 & 28 &$-1.41$ & 0.21 &  12.0 &$-$0.25&$-$0.34 &102.9 &103.0 
&  0.61 &$-$0.09 & $-$0.45~(1) \\
200613 & 25 & 18 &$-1.24$ & 0.18 &   ..  &   ..  &    ..  &103.5 &110.1 
&  0.64 &   0.06 &    0.01~(1) \\
\multicolumn{13}{c}{Subgiants}      \\
  1406 & 48 & 44 &$-1.39$ & 0.16 &  18.0 &  0.40 &   0.34 & 82.1 &105.7 
&  0.02 &   0.20 &    0.10~(2) \\
  1665 & 49 & 41 &$-1.34$ & 0.15 &   ..  &   ..  &    ..  & 86.7 &116.6 
&  0.10 &   0.10 & $-$0.13~(2) \\
  1445 & 35 & 33 &$-1.53$ & 0.13 &   ..  &   ..  &    ..  & 95.0 & ..   
&  0.19 &   0.06 &    0.33~(2) \\
  1563 & 44 & 38 &$-1.42$ & 0.17 &  17.4 &  0.38 &   0.33 & 97.0 &131.0 
&  0.25 &   0.11 &    0.17~(2) \\
  1400 & 70 & 41 &$-1.44$ & 0.14 &  15.3 &  0.33 &   0.28 &103.4 & ..   
&  0.29 &   0.10 &    0.34~(2) \\
  1461 & 25 & 32 &$-1.38$ & 0.24 &   ..  &   ..  &    ..  & 94.9 &144.4 
&  0.29 &   0.10 & $-$0.06~(2) \\
202063 & 29 & 38 &$-1.49$ & 0.27 &   ..  &   ..  &    ..  &120.4 &143.8 
&  0.44 &   0.13 &    0.33~(2) \\
  1481 & 55 & 46 &$-1.39$ & 0.15 &   ..  &   ..  &    ..  &122.5 &166.2 
&  0.54 &$-$0.07 &    0.69~(2) \\
  1460 & 38 & 32 &$-1.37$ & 0.15 &$<$14.0&$<$0.25&$<$0.19 &137.4 & ..   
&  0.65 &$-$0.19 &    0.86~(2) \\
\hline
\end{tabular}
\end{table*}

\section{ANALYSIS}

Equivalent widths (EWs) for a large number of lines were measured using an
authomatic procedure (similar to the one used in Bragaglia et al. 2001; full
details will be given elsewhere). Typical errors in the EWs are $\pm 3$~m\AA\
for NGC6397 stars, and $\pm 5$~m\AA\ for NGC6752 stars, but they are as low as
$\pm 2.5$~m\AA\ in the best cases (the Li lines at 6707~\AA). These values are
comparable with results obtained for much brighter stars at other telescopes.

One of the aims of the present analysis was to derive accurate reddening
estimates for GCs by comparing the observed colours with effective
temperatures $T_{\rm eff}$'s obtained from the spectra; for this reason our
temperatures are based on fitting of the wings of the H$_\alpha$\ profiles
with synthesized lines\footnote{We only used H$_\alpha$. $H_\beta$\ was not 
observed; the other members of the Balmer series were indeed observed, but
they are more contamined by metal lines, and their dependence on gravity, 
metal abundance and details in the convection much larger than for H$_\alpha$\
(see Fuhrmann et al. 1993)}. The model lines were computed according to the
precepts given in Castelli, Gratton \& Kurucz (1997); throughout this paper we
used model atmospheres extract by interpolation within the grid by Kurucz with
the overshooting option switched off (Kurucz 1993). Within these models,
convection is considered using a mixing length approach, with a pressure scale
height value of $l/H_p=1.25$. Internal errors of individual temperatures are
estimated to be 150-200 K for individual stars; these rather large errors are
mainly due to uncertainties in the flat fielding procedure, and are
typical for such temperature derivations when only one Balmer line is used
(see e.g. Fuhrmann, Axer \& Gehren 1994). The adopted temperature scale is
confirmed by analysis of line excitation: however, temperature from excitation
have slightly larger error bars due to the limited range of excitation for
lines measurable in warm, metal-poor dwarfs and subgiants.

For field stars, we compared these $T_{\rm eff}$'s (Column~7 of Table~1) with
those given by colours ($B-V$\ and $b-y$), using the calibration given by Kurucz
(1993); here we assumed that all these nearby (field) stars are unreddened.
Internal errors in $T_{\rm eff}$'s from colours are about 26 K. On average,
temperatures derived from H$_\alpha$\ are lower than those derived from colours
by $97\pm 38$~K, with a large  r.m.s scatter of 188 K. For field stars, we
then corrected the temperatures from colours to those derived from $H_\alpha$,
and averaged the two values, giving a weight 4 to temperatures from colours,
and 1 to temperatures derived from $H_\alpha$. These are our "best
temperatures" for the field stars (Column~8 of Table~1).

We may compare these "best temperatures" with those obtained by Alonso,
Arribas \& Martinez-Roger (1996) using the IR flux method (Column 9 of
Table~1). Excluding the discrepant case of HD132475, on average our best
$T_{\rm eff}$'s are lower than those given by Alonso et al. by $5\pm 14$~K
(r.m.s. residuals=45~K, 10 stars). Hence our $T_{\rm eff}$'s can be assumed to
be on the same scale of Alonso et al.

When plotted one on top of the other (separately: dwarfs and subgiants, and
stars in different clusters), the $H_\alpha$\ profiles for cluster stars look
undistinguishable. Furthermore, we were not able to find any correlation
between colours and temperatures either from line excitation or H$_\alpha$\
profiles. We then concluded that the stars are intrinsically very similar,
and that the slightly different values of colours and temperatures we
derived for each star are due to random errors. In the following analysis, we
have then adopted for all stars in these groups the same average temperatures
(note however that this assumption is not critical in the present discussion).

Surface gravities $\log g$\ were obtained from the location of the stars in
the colour magnitude diagram, assuming masses consistent with an age of 14~Gyr
(again, this assumption is not critical). As usual, microturbulent
velocities were obtained by eliminating trends of abundances derived from
individual Fe lines with expected line strength. 
Star-to-star scatter in Fe abundances within each group (same cluster, same
evolutionary phase) were reduced by adopting for all stars the same
microturbulent velocity; these average values were then adopted in the final
analysis.

The finally adopted atmospheric parameters (effective temperatures in
K/surface gravities/model metal abundances/microturbulent velocities in
km~s$^{-1}$) were as follows: NGC6397 TO-stars (6476/4.10/$-$2.04/1.32);
NGC6397 subgiants (5478/3.42/$-$2.04/1.32); NGC6752 TO-stars
(6226/4.28/$-$1.43/0.70); NGC6752 subgiants (5347/3.54/$-$1.43/1.10).

\begin{table}
\caption{Sensitivities of abundances to errors in the atmospheric parameters}
\begin{tabular}{lrrr}
\hline
Element & $\Delta T_{\rm eff}$ & $\Delta \log g$ & $\Delta v_t$ \\
        & +100~K               & +0.2~dex        & +0.2~km\, s$^{-1}$ \\
\hline
\multicolumn{4}{c}{TO-star} \\
$[$Fe/H$]$  &   +0.096 & $-$0.054 & $-$0.050 \\
$[$O/Fe$]$  & $-$0.161 &   +0.123 &   +0.048 \\
$[$Na/Fe$]$ & $-$0.037 &   +0.007 &   +0.032 \\
$[$Mg/Fe$]$ & $-$0.034 &   +0.006 &   +0.041 \\
$[$Al/Fe$]$ &   +0.001 & $-$0.012 &   +0.007 \\ 
\multicolumn{4}{c}{Subgiants} \\
$[$Fe/H$]$  &   +0.108 & $-$0.023 & $-$0.049 \\
$[$O/Fe$]$  & $-$0.200 &   +0.098 &   +0.044 \\
$[$Na/Fe$]$ & $-$0.044 & $-$0.010 &   +0.026 \\
$[$Mg/Fe$]$ & $-$0.055 &   +0.020 &   +0.043 \\
$[$Al/Fe$]$ & $-$0.080 &   +0.021 &   +0.038 \\ 
\hline
\end{tabular}
\end{table}

Sensitivities of abundances to errors in the atmospheric parameters are given
in Table~4. Errors in final abundances are mostly due to possible errors in
the adopted $T_{\rm eff}$'s. For our line list (dominated by low excitation
lines), Fe~I abundances raise by 0.10~dex, [O/Fe] values by -0.20~dex, and
[Na/Fe] ones by -0.04~dex for a 100~K increase in the adopted $T_{\rm eff}$.
Systematic errors in $T_{\rm eff}$'s are dominated by uncertainties in the
fitting of the $H_\alpha$\ profiles: they are about $\pm 90$~K for the dwarfs
and $\pm 60$~K for the subgiants, leading to errors in the [Fe/H] values of
$\pm 0.09$~dex and $\pm 0.06$~dex respectively for TO-stars and subgiants.
This value, appropriate for cluster stars, is smaller than that given for
field stars simply because our $T_{\rm eff}$'s for cluster stars are actually
the average over the values obtained for several stars, and errors for
individual stars are given by uncertainties in the flat fielding procedure,
that are only weakly affected by S/N. Corresponding errors in [O/Fe]'s are
$\pm 0.18$\ and $\pm 0.12$~dex; those in [Na/Fe] are $\pm 0.04$\ and $\pm
0.03$~dex. However, in the context of the O-Na anticorrelation, errors in
temperatures adopted for individual stars are more important. In the case of
NGC6397, a quite realistic estimate can be obtained by the star-to-star
scatter in Fe abundances, that is 0.032 dex, corresponding to an r.m.s. spread
of 33~K in the $T_{\rm eff}$'s. The adoption of a uniform temperature would
then yield errors of $\pm 0.07$~dex in [O/Fe]'s and $\pm 0.01$~dex in
[Na/Fe]'s (to be compared with the observed star-to-star scatter). In the case
of NGC6752, the star-to-star scatter of 0.096~dex in [Fe/H]'s also includes an
important contribution due to noise in the $EW$s: in fact the r.m.s. reduces
to 0.074~dex if only spectra with $S/N> $40 are considered. We conclude that
errors in individual $T_{\rm eff}$'s are $\lsim 73$~K. Related errors in
[O/Fe] and [Na/Fe] abundances are $\lsim 0.15$ and $\lsim 0.03$~dex. Again,
these are the values to be compared with the observed scatter.

\section{IRON ABUNDANCES}

Fe abundances were derived using a rather large number of lines, typically
$\sim 20$\ and $\sim 30-40$ for TO-stars and subgiants respectively. Full 
details about the line list, including $gf$'s, $EW$s and abundances from
individual lines, will be given in separate papers (in preparation).

The Fe abundance obtained for dwarfs in NGC6397 ([Fe/H]=$-2.02\pm 0.01$) is in
good agreement with that determined from subgiants ([Fe/H]=$-2.05\pm 0.03$).
The average of the two groups is [Fe/H]=$-2.03\pm 0.02$\ (this is the internal
error of our analysis; systematic errors are $\pm 0.04$~dex). This value is
less than that derived from giants in Carretta \& Gratton (1997:
[Fe/H]=$-1.82\pm 0.04$), and Zinn \& West (1984: [Fe/H]=$-1.91$). However it
agrees very well with the value obtained by Minniti et al. (1993:
[Fe/H]=$-1.96\pm 0.04$), and the recent, comprehensive analysis of giants and
subgiants by Castilho et al. (2000: $-2.00\pm 0.05$).

The equilibrium of ionization is not well reproduced: abundances from neutral
Fe lines are 0.11 dex larger than those from singly ionized Fe lines. The same
result is obtained for NGC6752 and our field stars. Note that a smaller
difference (0.07~dex) in the same sense is also present in our solar reference
analysis (in that case we obtained $\log n({\rm Fe})=7.52$\ from Fe~I lines,
and 7.45 from Fe~II lines using the solar model atmosphere from Kurucz 1993,
with no overshooting). We then think most of the difference for the Sun is due
to either the adopted $gf's$\ values (these are laboratory values taken from
recent literature) or to the model atmospheres (models might underestimate the
temperature gradient in real atmospheres, perhaps due to an inappropriate
consideration of convection). The residual difference for the program stars
might be due to the adoption of slightly too high $T_{\rm eff}$'s ($\sim
40-50$~K) or too low gravities (by $\sim 0.1$~dex). Note that it cannot be due
to departures from LTE, because the expected dominating effect
(overionization: Idiart \& Th\'evenin 1999; Gratton et al. 1999) would lead to
larger abundances from Fe II lines than from Fe I ones (opposite to
observations).

The star-to-star scatter in [Fe/H] values is extremely small: the r.m.s.
scatter is only 0.04 dex (i.e. 10\%) for NGC 6397. This  seems a very
homogeneous cluster as far as Fe abundances are considered.

On the other side, the [Fe/H] value for NGC6752 ([Fe/H]=$-1.42\pm 0.02$,
internal error; systematic error is again $\pm 0.04$~dex), obtained both from
TO  and subgiant stars, which agree completely, coincides with what derived
from giants by Carretta \& Gratton (1997: [Fe/H]=$-1.42\pm 0.02$); it is
somewhat larger than the value quoted by Zinn \& West ([Fe/H]=$-1.54$) and
Minniti et al. (1993: [Fe/H]=$-1.58\pm 0.04$). The spectra of NGC6752 have a
S/N lower than those in NGC6397, since we chose to observe more stars, even at
a lower S/N. The scatter of abundances for individual lines (from 0.12 to 0.27
dex) is larger than that obtained for stars in NGC6397, roughly in agreement
with the lower S/N.


For field stars we may compare the present Fe abundances with those derived by
Carretta, Gratton \& Sneden (2000). Limiting ourselves to only those stars for
which Carretta et al. considered high dispersion abundances, the present Fe
abundances are smaller on average by $-0.05\pm 0.02$~dex (11 stars,
r.m.s.=0.08~dex). The slightly lower metal abundances are due to lower $T_{\rm
eff}$'s adopted in the present paper.

\section{THE O-NA AND MG-AL ANTICORRELATION}

In our analysis, Oxygen abundances were derived from the permitted infrared
triplet (7771-74~\AA): the forbidden lines are too weak in the program stars
to be reliably measured, and we did not observe the region of the OH band in
the near UV because the required exposure times would be prohibitive even for
an 8~meter telescope. Na abundances are based on the quite strong doublet at
8184-90~\AA, clearly visible in all stars. We checked that telluric lines were
not blended with the stellar features. For both O and Na, our abundances
included non-LTE corrections, computed following the precepts of Gratton et
al. (1999): however, these corrections are small and they do not affect any of
the conclusion reached in this paper. Mg abundances are based on a few high
excitation lines in the blue and yellow portion of the spectrum (typically
three to four lines were measured for each star). Finally, whenever possible
Al abundances were measured using the high excitation IR doublet at 
8772.9/73.9~\AA. However, this was possible only for subgiants in NGC6752,
since in the other cases this doublet is too weak, and the only chance to
measure Al abundances is by using the resonance doublet at 3944/61~\AA.
However, this lines, while stronger, are not ideal abundance indicators due to
saturation and the presence of large departures from LTE (see e.g. Fran\c cois
1986; Beyley \& Cottrell 1987; and the discussion in Gratton \& Sneden 1988).
We have not prepared yet an adequate model for the Al atom to account for such
departures from LTE; anyway abundances obtained by an LTE analysis of these
lines may still be useful to discuss the Mg-Al anticorrelation insofar only
stars with very similar atmospheric parameters are considered, because in this
case we expect that departures from LTE should be essentially the same for all
such stars.

For NGC6397, the O abundance is [O/Fe]=$+0.21\pm 0.05$\ (internal error;
systematic errors are $\sim 0.1$~dex), and the average [Na/Fe] ratio is
[Na/Fe]=$+0.20\pm 0.05$. This O excess is quite small in comparison to the
values usually found for metal-poor stars (see e.g. Gratton et al. 2000), but
it agrees very well with the mean values determined by Minniti et al. (1997:
[O/Fe]=+0.19) and Norris \& Da Costa (1995: [O/Fe]$\sim 0.1$) from analysis of
the forbidden lines in red giant spectra. Also the Mg abundance we get is quite
low: this point will be discussed in a next paper. On the other side, the
[Na/Fe] value we found agrees well with that determined for a single red giant
by Carretta (1994: [Na/Fe]=+0.22), and for two more by Norris \& Da Costa
(1995: [Na/Fe]$\sim 0.2$), while it is somewhat larger than the value of
[Na/Fe]=$-0.01$\ found by Minniti et al. (1997). We wish to remark that none
of the program stars seem to be oxygen-poor and sodium-rich. The star-to-star
scatter in our determinations (0.15 and 0.14 dex, r.m.s., respectively) is
larger than expected from errors in the EWs alone (we would expect r.m.s.
values of $\sim 0.10$~dex): however it may be justified by some star-to-star
scatter in the atmospheric parameters (within the measuring errors), not
accounted for in our analysis (we are assuming that all dwarfs and subgiants
may be analyzed using the same model atmospheres), and by a small offset in
the results for dwarfs and subgiants (again, likely due to small errors in the
adopted set of atmospheric parameters, within our quoted uncertainties). Given
the small spread in Fe abundances, and the absence of a clear O-Na
anticorrelation, we conclude that NGC6397 is indeed a very homogenous cluster.
On the whole, our results indicate that stars in this cluster conforms to the
paradigma set by field stars, confirming earlier findings by Bell \& Dickens
(1980).

\begin{figure*} 
\psfig{figure=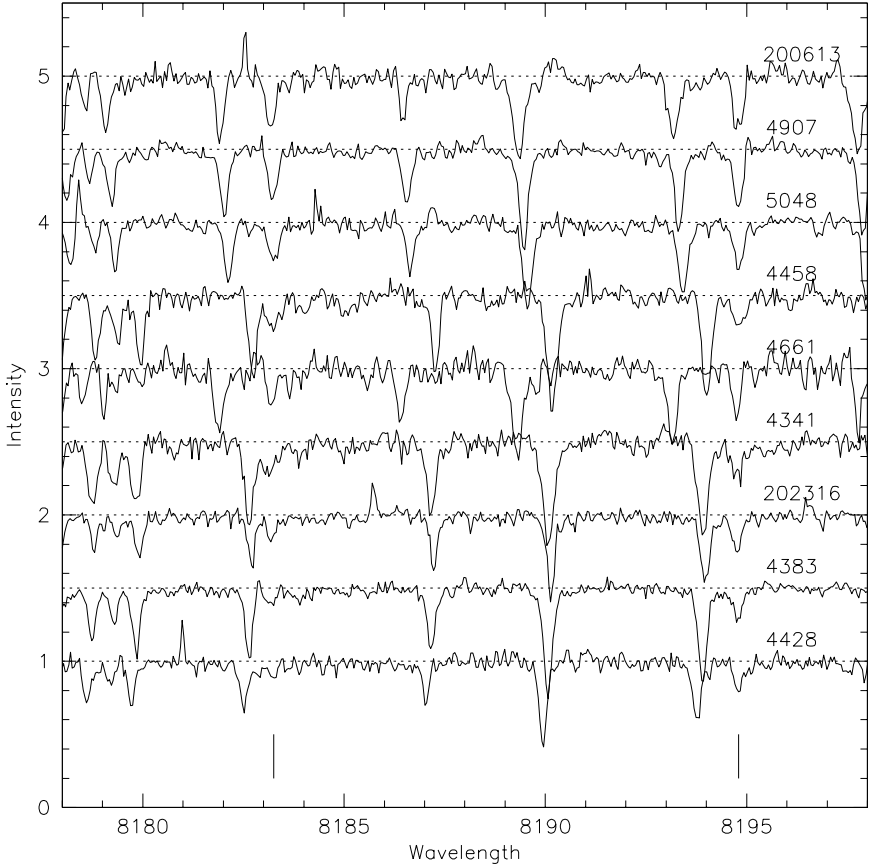,width=13.0cm,clip=}
\caption[]{ Spectral region including the 8184-90~\AA ~Na I doublet in NGC6752
TO-stars (stars are ordered according to decreasing Na abundances). The position
of the Na lines is marked. Note that all these stars essentially have the same
temperature, gravity, overall metal abundance and microturbulent velocity, so
that observed variations in the line strengths can be directly interpreted as
spread in the abundances}
\label{f:fig3} 
\end{figure*}

\begin{figure*} 
\psfig{figure=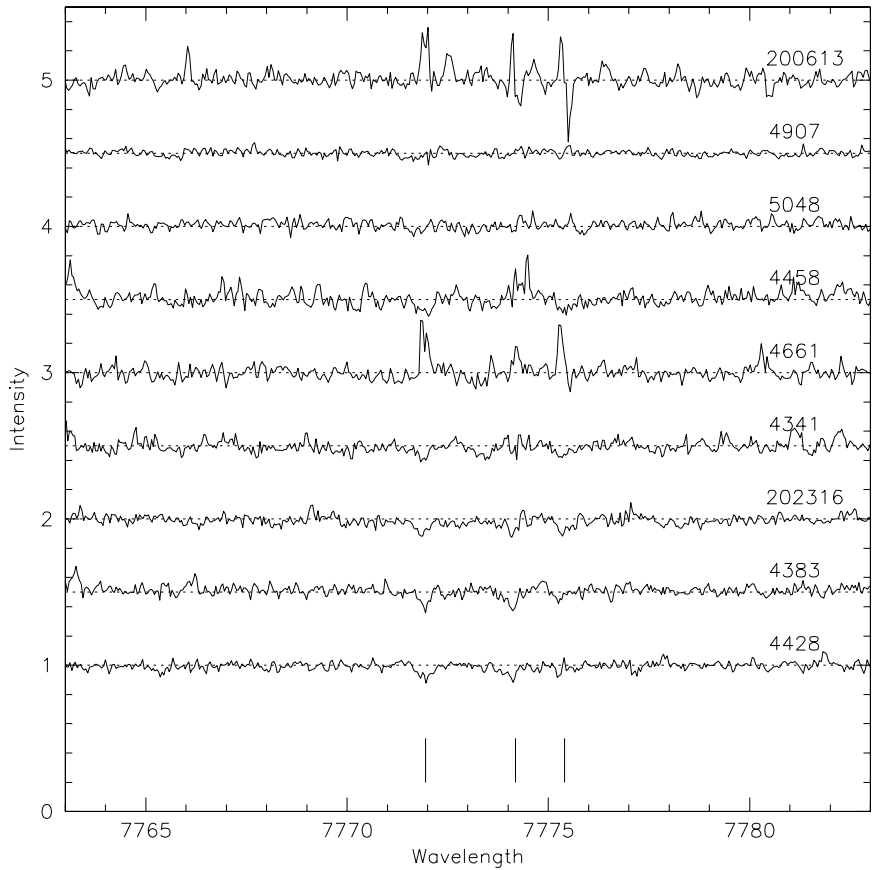,width=13.0cm,clip=}
\caption[]{ 
Same as Figure 3, but for the region including the 7771-74~\AA ~OI triplet 
in NGC6752 TO-stars (stars are ordered according to decreasing Na abundances).
The position of the O lines is marked.}
\label{f:fig4} 
\end{figure*}  

\begin{figure} 
\psfig{figure=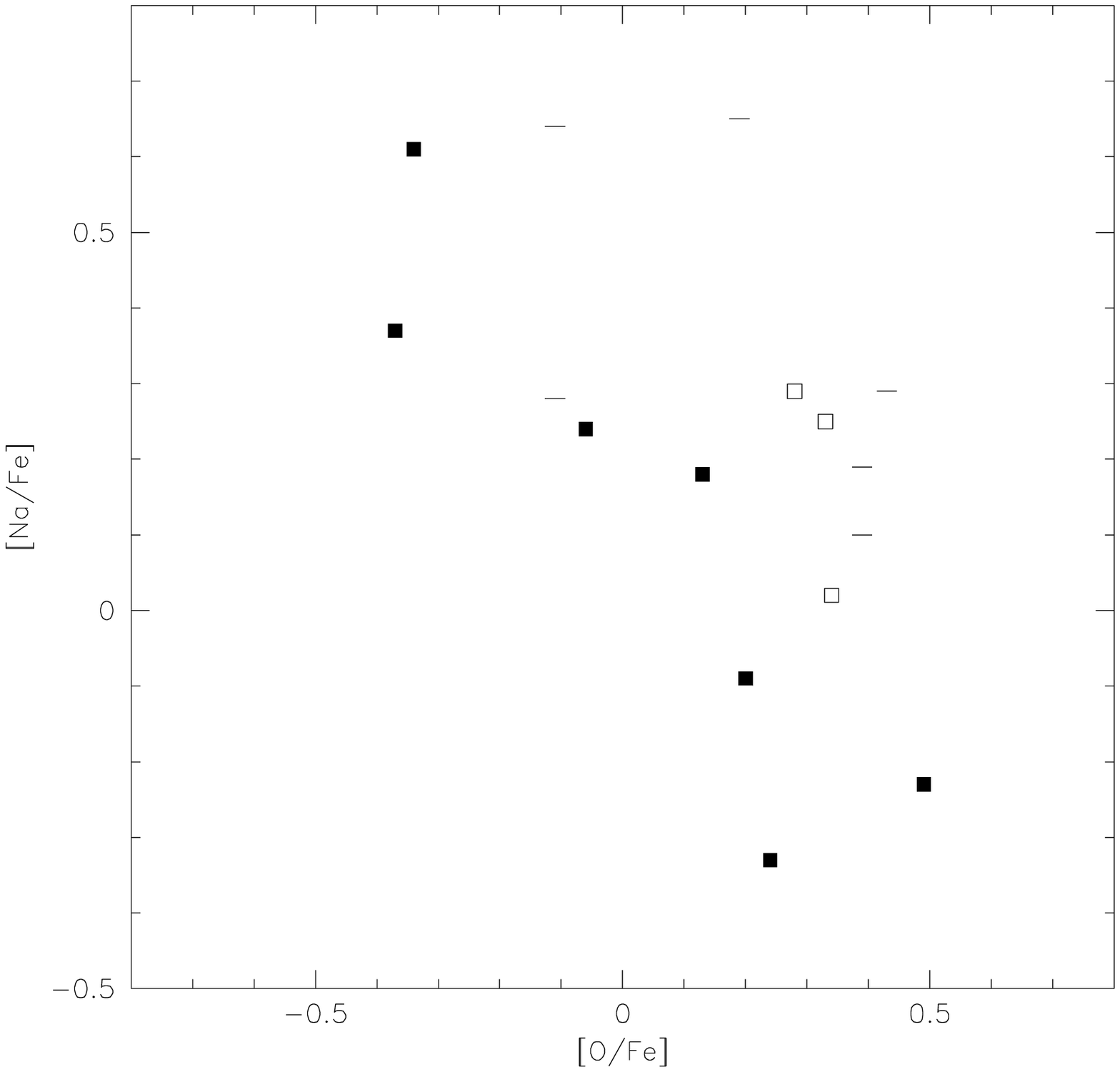,width=8.8cm,clip=}
\caption[]{ Run of the abundances of Na and of O for stars 
in NGC6752. Filled squares are TO-stars, open squares are subgiants, and lines
represent upper limits (for O) } 
\label{f:fig5} 
\end{figure}  

NGC6752 presents a very different scenario (this is not too surprising, in
view of the variations of the strength of CN and CH bands among stars on the
MS already noticed by Suntzeff \& Smith 1991). There is a clear
anti-correlation between O and Na abundances even for stars at the TO. This is
clearly illustrated by Figures~\ref{f:fig3} and \ref{f:fig4}, where we plotted
the spectral regions including the Na and O lines in these stars (stars are
ordered according to decreasing Na abundances). Notice that all these stars
have the same (or at least very similar) temperature, gravity, overall metal
abundance and microturbulent velocity, so that observed variations in the line
strengths may be directly interpreted as spread in the abundances: then, in
spite of the fact that a few of the spectra are somewhat noisy, this result
does not depend on details of the analysis, but rather is a solid purely
observational evidence, largely independent of all the assumptions made. We
found that Na abundances span a range of almost an order of magnitude. O
abundances also change by a rather large factor; while not as extreme as in
M13, we found a rather extensive O-Na anticorrelation in NGC6752 too. This
anti-correlation is shown in Figure~\ref{f:fig5}. There is a small offset in
the O-Na anticorrelation between TO-stars and subgiants, but we think this is
an artefact of the analysis (slightly incorrect offsets in temperatures and
gravities).

\begin{figure} 
\psfig{figure=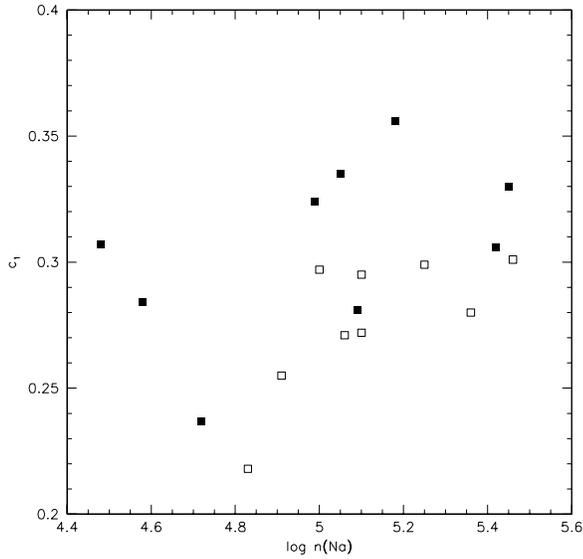,width=8.8cm,clip=}
\caption[]{ Run of the $c_1$\ index from Str\"omgren photometry with the Na
abundance for stars in NGC6752. Filled squares are TO-stars, and
open squares are subgiants }
\label{f:fig6} 
\end{figure}  

There is some correlation between $c_1$\ and the O-Na trends; hence, as
expected, this can be used to select candidates. Note however that the
scatter in this relation is quite large (see Figure~\ref{f:fig6}).

\begin{figure} 
\psfig{figure=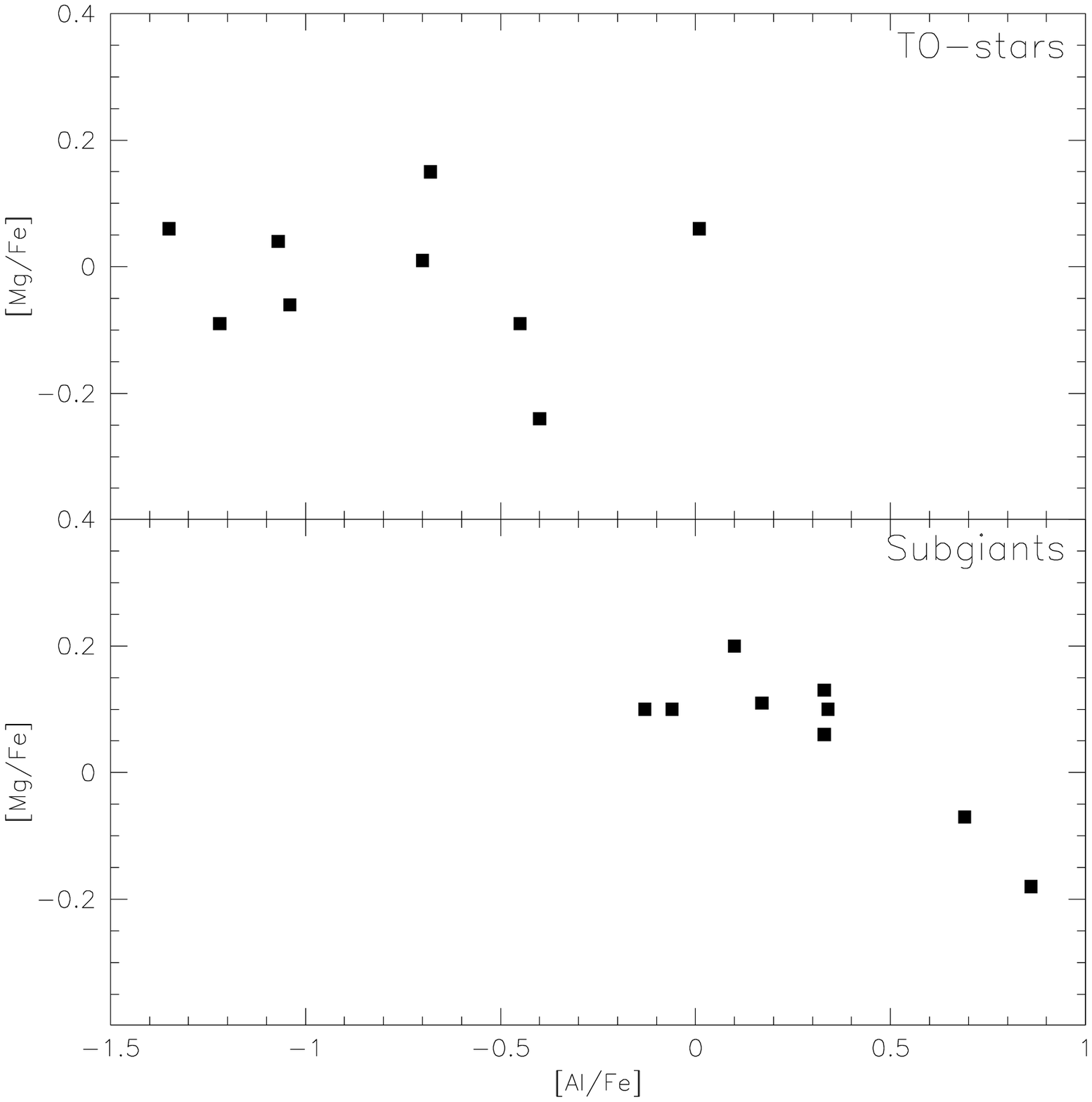,width=8.8cm,clip=}
\caption[]{ Run of the abundances of Mg against that of Al for stars in
NGC6752. The upper panel shows results for TO-stars; the lower panel those for
subgiants. Note that Al abundances for TO-stars were obtained using the
resonance doublet, that is affected by large departures from LTE. This causes
the large offset of Al abundances obtained by the two groups of stars}
\label{f:fig7} 
\end{figure}  

Figure~\ref{f:fig7} displays results for Mg and Al in NGC6752. The
star-to-star variations in the Al abundances are large: $\sim 1$~dex among
subgiants, and even more among TO-stars. These large variations cannot be
explained by observational errors, and (for TO-stars) by any plausible
differential non-LTE effect among stars with very similar atmospheric
parameters. On the other side, star-to-star variations in the Mg abundances
are much smaller ($\lsim 0.4$~dex), and they may be seen clearly only among
subgiants; for TO-stars, the used Mg lines are quite weak and a larger
scatter in the abundances is due to errors in the $EW$s. Data for subgiants,
for which reliable Al abundances could be obtained from the high excitation Al
I doublet at 8772.9/73.9~\AA, indicate that also for these elements there is a
clear anticorrelation between Mg and Al. Stars 1460 and 1481 (the two most
Na-rich subgiants) are clearly Mg-poor, Al-rich. We underline here that while
the Mg-Al anticorrelation is not obvious in panel a of Figure~\ref{f:fig7}, we
think that such an anticorrelation exists also for TO-star because the Al
abundances show a very large star-to-star variation, correlated with Na
abundance variations; we think the anticorrelation is less obvious in this
case because we had to measure the resonance doublet of Al at 3944/61~\AA, not
an ideal abundance indicator as we mentioned above, and moreover Mg abundances
have larger errors, comparable to the star-to-star variations seen among
subgiants.

We are now preparing line lists to study CH and CN abundances. However, it
seems clear that the strength of these bands is correlated with Na and O
abundances. Also, there is a correlation with the Stromgren $c_1$\ index,
albeit the scatter is quite large. Finally, in a forthcoming paper we will
present a full discussion of the Li abundances in these stars. Here we
anticipate that Li abundances in NGC6752 stars seem also anticorrelated with
the Na ones.

\section{SUMMARY AND CONCLUSIONS}

We have presented first results of the ESO Large Program 165.L-0263: we have
used the UVES spectrograph on VLT2 to obtain high resolution ($R\gsim 40,000$)
spectra for a quite large number of stars at the turn-off (14 stars between
the two GCs) and the base of the subgiant branch (12 stars) in the globular
clusters NGC6397 and NGC6752. Thanks to the efficiency and large spectral
coverage of UVES we were able to obtain reliable $EW$s for a number of lines
of Fe, Li, O, Na, and other elements. The main results of this first analysis
are:

(i) The [Fe/H] value for NGC6397 is [Fe/H]=$-2.03\pm 0.02\pm 0.04$\ (internal 
and systematic errors), less
than derived from giants by Carretta \& Gratton (1997) and Zinn \& West
(1984), but in agreement with the recent analysis of giants and subgiants by
Castilho et al. (2000). The [Fe/H] value for NGC6752 ([Fe/H]=$-1.42\pm 0.02\pm
0.04$) coincides with the value derived from giants by Carretta \& Gratton
(1997), and is slightly higher than the value quoted by Zinn \& West (1984)

(ii) In both clusters, [Fe/H] obtained for TO stars agrees perfectly (within a
few percents) with that obtained for stars at the base of the giant branch;
this is a constraint on the impact of diffusion in stellar models. The
star-to-star scatter is extremely small: the r.m.s. scatter is only 0.04 dex
(i.e. 10\%) for NGC 6397. This  seems a very homogeneous cluster as far as Fe
abundances are considered.

(iii) For NGC6397, the O abundance is [O/Fe]=$+0.21\pm 0.05$. This is a quite
low value in comparison with those usually found in metal-poor stars, but it
agrees well with previous determinations for red giants. The scatter of
individual values is small, and none of the program stars seem to be
oxygen-poor.

(iv) For NGC6752, there is a clear anticorrelation between Na and O among TO
stars and subgiants (similar to that seen among giants in this and other
clusters). Also, an anticorrelation is observed between Mg and Al, most
clearly among subgiants, but likely existing also among TO-stars. Na-poor
stars (i.e. stars with [Na/Fe]$<0.2$) in NGC6752 have [O/Fe]=+$0.30\pm 0.04$\
(7 stars, r.m.s.=0.10 dex). Extremes in Al abundances differ by over 1 dex,
while for Mg the star-to-star scatter is smaller. The large variations in Mg
and Al abundances suggests that nearly half of Mg has been converted
into Al in the most Al-rich stars (1481 and 1460); note that these
stars are also the most Na-rich ones. Note that given the adopted sample 
selection criteria, extreme cases of Na-poor and Na-rich stars may be 
overrepresented among observed stars.

We think the present results are very difficult to be reconciled with deep
mixing scenarios. To our knowledge, there are not appropriate calculations for
main sequence stars. For giants, some calculations have been made by  Langer,
Hoffman \& Sneden (1993) and Denissenkov \& Tout (2000): they show that the
temperature required for p-capture on $^{24}$Mg to finally produce the Mg-Al
anticorrelation is $\sim 6\times 10^7$~K; even if Al is produced starting from
the far less abundant $^{25}$Mg isotope, the temperature
required is $\gsim 4\times 10^7$~K; finally, the temperature required for
extensive O-burning and Na production by p-capture on $^{22}$Ne is $\sim
3\times 10^7$. All these values are much higher than expected even at the
center of a globular cluster TO-star ($\sim 2\times 10^7$~K). Admittedly,
these computations assume densities about 20 times lower than expected at the
center of TO-stars, and much shorter timescales than MS lifetimes. However
complete mixing of MS stars is unacceptable for several other reasons (they
would e.g. bring large amount of fresh H to the center); furthermore, Li would
be completely destroyed (while we see some Li even in O-depleted stars; paper
in preparation). Hence, we think deep mixing scenarios cannot explain our
results.

We are then forced to some primordial mechanism, like those proposed years ago
by Cottrell \& Da Costa (1981) and D'Antona, Gratton \& Chieffi (1983). In
both scenarios, the inhomogeneities are due to the mass lost by intermediate
mass stars ($M = 4 - 5M_\odot$) during the Asymptotic Giant Branch (AGB)
evolution and the planetary nebula expulsion: the two scenarios differ because
Cottrell \& Da Costa think of a prolonged star formation, with most recently
formed stars having a different chemical composition from the first ones; while
D'Antona et al. consider pollution of the outer layers of already formed stars
by other objects in the cluster. Anyhow, in both scenarios the intracluster
gas is heavily nuclearly processed, due both to the occurrence of the third
dredge-up from the helium buffer (Iben 1975) and by Hot Bottom Burning (HBB)
at the basis of the convective envelopes of these massive AGB stars. Models
for these evolutionary phases (Sackmann and Boothroyd 1992, Ventura et al.
2000) successfully explain the occurrence and evolution of the Lithium rich,
Oxygen rich massive AGBs in the Magellanic Clouds (Smith et al. et al. 1995).
Very recent models by Ventura et al. (2001) show that, in full stellar models
computed for these intermediate mass stars at the low metallicities of
Globular Clusters, the HBB temperature can reach values as large as $10^8$K.
At this temperature, the complete CNO cycle operates, depleting oxygen. At the
same time, p-captures on $^{24}$Mg and $^{20}$Ne produce Al (see also
Denissenkov et al. 1998) and Na. This model can then explain the
anticorrelations O-Na and Mg-Al. In this context it is very interesting to
note that no O-Na anticorrelation is seen in NGC6397. Also, Li abundances in
this cluster follows the paradigma set by field metal-poor stars (Castilho et
al. 2000). NGC6397 is a quite small cluster (mass $\lsim 10^5~M_\odot$, from
the integrated magnitude $M_V=-6.58$, Harris 1996, and a mass-to-light ratio
of $\sim 2$, typical for a globular cluster). On the other side, the O-Na
anticorrelation is seen in the more massive cluster NGC6752 (mass $\sim
2\,10^5~M_\odot$, from the integrated magnitude $M_V=-7.68$, Harris 1996, and
the same mass-to-light ratio used for NGC6397). A (cluster) mass threshold
should be present in accretion scenarios (see e.g. Gratton 2001), and likely
also in the prolonged star formation models. We plan to address thoroughly
such problems in forthcoming papers.

\begin{acknowledgements}
{ This research has made use of the SIMBAD data base, operated at CDS,
Strasbourg, France. We wish to thank V. Hill for help during the observations,
and P. Bertelli for useful comments}
\end{acknowledgements}

\end{document}